\documentclass[fleqn,twoside]{article}

\usepackage[headings]{espcrc2}

\readRCS
$Id: espcrc2.tex,v 1.2 2004/02/24 11:22:11 spepping Exp $
\usepackage{graphicx}
\usepackage[figuresright]{rotating}

\newcommand{\AmS}{{\protect\the\textfont2
 A\kern-.1667em\lower.5ex\hbox{M}\kern-.125emS}}
\usepackage{amssymb}
\usepackage{epsfig}
\newcommand{\slsh}[1]{\mbox{$\not\! #1$}}
\newcommand{\bi}[1]{\bibitem{#1}}
\newcommand{\bm}[1]{\mbox{\boldmath $#1$}}

\newcommand{\FF}{{\mathcal F}}

\newcommand{\bea}{\begin{eqnarray}}
\newcommand{\eea}{\end{eqnarray}}

\newcommand{\nn}{\nonumber \\}
\newcommand{\nnn}{\nonumber}
\newcommand{\be}{\begin{eqnarray}}
\newcommand{\ba}{\begin{array}}
\newcommand{\ea}{\end{array}}
\newcommand{\ee}{\end{eqnarray}}

\newcommand{\Tr}{{\rm Tr}}

\newcommand{\eg}{{\it e.g.}\ }

\title{$T$-Odd Effects in Unpolarized Drell-Yan Scattering}

\author{Leonard P. Gamberg\address[psu]{Division of Science,
Penn State-Berks Lehigh Valley College, \\
         Reading, PA 19610, USA}
       \thanks{Email Address: lpg10@psu.edu},
       Gary R. Goldstein\address[tu]{Department of Physics and
                Astronomy, Tufts University,  Medford, MA 02155, USA}
       \thanks{Email Address: gary.goldsetein@tufts.edu}}

\begin{document}

\begin{abstract}
We consider the leading
twist $T$-odd  contributions as  the dominant source of the
$\cos 2\phi$ azimuthal asymmetry in unpolarized
$p\bar{p}\rightarrow \mu\mu^+\ X$
dilepton production in Drell-Yan Scattering. This asymmetry contains
information on the distribution of quark transverse spin
in an unpolarized proton.
In a parton-spectator framework we estimate 
these asymmetries at $50\ {\rm GeV}$ 
 center of mass energy. This azimuthal asymmetry  
is interesting in light of proposed 
experiments at GSI, where an anti-proton beam is ideal for studying the
transversity properties of quarks due to the dominance of {\em valence} quark
effects.
\vspace{1pc}
\end{abstract}

\maketitle

\section{Introduction}
\vspace*{-0.25cm}
One of the persistent challenges confronting the QCD
parton model is to provide a theoretical basis to understand
the experimentally significant azimuthal and transverse
spin asymmetries that emerge in exclusive, inclusive and semi-inclusive
processes~\cite{heller,e615,na10,E704,star,hermes}.
Generally speaking, the spin dependent amplitudes for the scattering will
contribute to non-zero transverse single spin asymmetries (SSA) if there are imaginary parts of
bilinear products of those amplitudes that have overall helicity change. For
two-body exclusive reactions, SSA requires there to be an imaginary part of the product of an
helicity non-flip with an helicity flip amplitude. For inclusive reactions, the same conclusion can be
reached by taking the amplitudes as two-body helicity amplitudes for the production of a fixed hadron
and a state $|X>$. Through the generalized optical theorem, SSA in inclusive reactions can be related to
discontinuities in helicity flip three-body forward scattering amplitudes~\cite{gando}. That
is essentially spin kinematics. Dynamically there must be quantum field theory contributions to the
relevant amplitudes. In perturbative QCD (PQCD), applicable to the hard scattering region, to obtain
an imaginary contribution to quark and/or gluon scattering processes demands introducing higher order
corrections to tree level processes. One approach incorporates the requisite phases
through interference of tree level and one-loop contributions in PQCD in an
attempt to explain spin asymmetry in $\Lambda$ production~\cite{gandd}. On
general grounds  the helicity conservation
property of massless QCD predicts that such contributions are  small,
going like $\alpha_s m/Q$, where $\alpha_s$ is the strong coupling, $m$ represents a non-zero quark mass
and $Q$ represents the hard QCD scale~\cite{kane,gandd}. Such contributions have failed
to account for the large SSA observed in $\Lambda$ production~\cite{heller}. On the other hand,  
the twist three quark-quark and
quark-gluon correlations described in~\cite{et}
and~\cite{qs} hold promise to describe the phenomena at large $p_T$. 

However, considering the soft contributions to hadronic
processes opens up the possibility that there are
non-trivial transversity parton distributions that can
contribute to transverse spin asymmetries.  Ralston and
Soper introduced~\cite{RS} the now well known chiral-odd transversity
transfer distribution function~\cite{artru,rat} $h_1(x)$
which can play a role in doubly polarized Drell-Yan  processes~\cite{RHIC}.
$h_1(x)$ can also be measured in semi-inclusive deep
inelastic scattering (SIDIS)~\cite{colnpb}. For SSA in SIDIS
transverse momentum must be acquired to lead to
appropriate helicity changes. In describing transverse asymmetries
this is particularly important when
the transverse momentum is sensitive to intrinsic quark momenta.  Here the
effects are associated with non-perturbative transverse
momentum distribution functions~\cite{soper} (TMD),
where transverse single spin
asymmetries indicate  so called $T$-odd correlations between
transverse spin and longitudinal and intrinsic quark transverse momentum.
The  $T$-odd distributions~\cite{sivers,ansel,boermul}
are  of importance as they possess both transversity properties and the
necessary phases to account for SSA and azimuthal 
asymmetries~\cite{colnpb,boerich,bhs}.
Formally, these phases can be generated from the
gauge invariant definitions of the $T$-odd
quark distribution functions~\cite{colplb,jiyuan,gg,boerpij}.
Asymmetries that involve $T$-odd
TMD and fragmentation functions
are indicative of
a rich set of correlations among
transverse momenta of quarks and/or hadrons and the
transverse spin of the reacting  hadrons and/or quarks.  
In contrast to the  SSA generated from the interference of
tree-level and one loop correction in PQCD, 
such effects go like $\alpha_s {<}k_\perp {>}/M$,
where now $M$ plays the role of the chiral symmetry breaking scale and
$k_\perp$ is characteristic of quark intrinsic motion.

$T$-odd distributions only exist by virtue of non-zero parton
transverse momenta~\cite{ansel,tangmul,boermul}.
They also correspond to distributions that would
vanish at {\em tree level} in any $T$--conserving model of  hadrons and
quarks. In this sense they are similar to the decay amplitudes for
hadrons that involve single spin asymmetries which are non-zero due to
final (and/or initial) state strong interactions.  Their existence was
suggested by Sivers to account for the significant SSA in
inclusive reactions (\eg $p\, p^\uparrow\rightarrow\, \pi X$)
~\cite{sivers,ansel}, by Collins in SIDIS~\cite{colnpb}, and by Boer~\cite{boerich} in 
Drell-Yan scattering.

A great deal of progress has been made in 
the last several years, following the realization that this link goes
further, as Brodsky, Hwang and Schmidt first showed~\cite{bhs} and
several researchers generalized thereafter~\cite{jiyuan,gg}.
Detailed model  calculations of these functions have been performed in parton inspired 
spectator-models of quark-hadron interactions where absorptive effects
are generated by interference between gluon loop corrections to tree-level 
 TMDs. These loop calculations were applied to SIDIS~\cite{bhs,jiyuan,metz,gg,ggoprh,ggoprf,bac,bacmetz} 
and  Drell-Yan processes~\cite{bbh,ggmar,ma}, thereby giving rise to 
predictions  for SSA and azimuthal asymmetries.

The  importance of quark distribution functions 
 was recognized some 35 years ago by Drell
and Yan~\cite{dy} when they considered  high energy hadron scattering that produces large invariant mass 
lepton pairs as a fundamental 
probe of quark-antiquark distribution functions. Furthermore, considering various asymmetries and
polarization  phenomena in Drell-Yan
processes can uncover relevant products of spin dependent distributions~\cite{boerich}.
Indeed,  Drell Yan $p\bar{p}$ scattering is a preferred reaction to study the 
the role that $T$-odd quark distribution functions 
play in the transverse  spin structure of the proton
through spin and azimuthal asymmetries in QCD~\cite{boerich}.   
That is the direction we pursue herein.
\vspace*{-0.25cm}
\section{Drell-Yan and $T$-Odd Correlations}
\vspace*{-0.15cm}
At the parton level the Drell-Yan cross section will receive contributions from quark-antiquark 
annihilation into the heavy photon.  In unpolarized Drell-Yan scattering
early cross section data as a function of the transverse momentum of the
muon pair indicated deviations from the Bjorken scaling prediction~\cite{cfs,cerndy} .
The implication was that the collinear approximation was insufficient to describe 
the data~\cite{altgrec,css85}.
Transverse momentum of a parton arises due to hard Bremsstrahlung of gluons,
which is calculable from PQCD when the momentum transfers are
large~\cite{col79}.
On the other hand, quark confinement implies that quarks have soft,
primordial or intrinsic transverse momenta $\bm{k}_\perp$. This latter effect is
significant at low transverse momentum, $\bm{q}_T\ll Q$.
$\bm{q}_T$ dependence has been incorporated into
the factorized Drell-Yan model~\cite{RS,cs} by extending the parton probability
distribution to be a function of $\bm{k}_\perp$~\cite{soper}~\footnote{
New work on the factorization theorem for Drell-Yan can be found in~\cite{jima}.}
\be
\int d\bm{k}_\perp{\mathcal P}(k_\perp,x)=f(x)\,  .
\ee
If the parton distributions within the incoming hadrons have transverse momentum dependence, there will be a 
continuum of values of their $k_\perp$ for which a time-like photon of fixed 4-momentum will be formed. 
Ignoring or summing over spin (and the lepton pair orientation),
the $k_\perp$ dependent distribution functions appear in the differential cross section,
\bea
\frac{d\sigma}{dq^2 dy d^2q_T}&=&
\frac{4\pi\alpha^2}{3Q^4}\sum_a e^2_a
\int d^2k_{\perp}\int d^2p_{\perp}
\nn && \hskip -2.5cm \delta^{(2)}
\left(\bm{k}_\perp+\bm{p}_\perp-\bm{q}_T  \right)
f_{a/A}(x,k_\perp)\bar{f}_{\bar{a}/B}(\bar{x}, p_\perp)
\nnn
\eea
where  $f_{a/A}(x_1,k_\perp)$ is a distribution function for a quark $a$ to be found in
hadron $A$ with transverse momentum $\bm{k}_\perp$ and longitudinal momentum fraction $x_1$
and $\bar{f}$ is the corresponding anti-quark distribution in hadron $B$.

Once transverse momentum dependence of parton distributions enters
the picture of scattering processes a much larger set of 
transverse momentum  distribution (TMD) 
and fragmentation functions\cite{tangmul}
become possible and relevant, particularly for spin asymmetries.
Among such functions are the possible leading twist
$T$-odd quark distribution~\cite{sivers} and fragmentation
functions~\cite{colnpb}. 

In SIDIS with an unpolarized target, an expectation value of
$i\bm{\mathsf s}_T\cdot\left({\bm P}\times {\bm k}_\perp\right)$,
indicates  a $T$-odd correlation of transverse
quark polarization with the proton's momentum and the
intrinsic quark transverse momentum in an unpolarized
nucleon  while  $i\bm{\mathsf s }_T\cdot(\bm{p}\times \bm{P}_{h\perp})$,
corresponds to that of a fragmenting quark's polarization
with quark and transverse pion momentum, $\bm{P}_{h\perp}$.
These correlations enter the {\em unpolarized}
cross-section through convolutions
with $h_1^\perp$~\cite{boermul} and the Collins
fragmentation function $H_1^\perp$~\cite{colnpb}.
The resulting $\cos 2\phi$  asymmetry is not suppressed
by $1/Q$, where $Q$ represents
the scale in SIDIS - the
$T$-odd contribution is at leading twist~\cite{boermul,ggoprh,ggoprf}. An analogous unpolarized double 
$T$-odd azimuthal asymmetry enters the 
Drell-Yan process~\cite{boerich}.  For the Drell-Yan process the angular dependence can be expressed as
\bea
\frac{d N}{d\Omega}\hspace*{-.25cm}&\equiv &\hspace*{-.25cm}\left(\frac{d\sigma}{dQ^2 dy d\bm{q}_T^2}\right)^{-1}
\frac{d\sigma}{dQ^2 dy d\bm{q}^2_T d\Omega}
\nn \hspace*{-.25cm}&=&\hspace*{-.25cm}  
\frac{3}{4\pi}\frac{1}
{\lambda+3}\Big( 1+\lambda\cos^2\theta \hspace{-.05cm}+\hspace{-.05cm}\mu \sin^2\theta \cos\phi
\nn && \hspace{2.5cm}+\frac{\nu}{2} \sin^2\theta \cos 2\phi\Big).
\label{cross}
\eea
The solid angle $\Omega$ refers to the lepton pair orientation in the pair rest frame relative to the boost direction and the incoming hadrons' plane~\cite{cs}.  $\lambda, \mu, \nu$ are functions that depend 
on  $s, x, m_{\mu\mu}^2, \bm{q}_T$,
the center of mass energy, the fraction of quark momentum in the hadron ,
the invariant mass of the produced lepton pair, and the transverse
momentum of the dimuon pair.~\footnote{We are
working in the Collins-Soper frame where $\bm{q}_T$ retains its meaning.}

By keeping the $\Omega$ dependence in the convolution,
Collins and Soper~\cite{cs} could project out the photon angular dependence in Eq.~(\ref{cross})
and obtain, among the other asymmetry functions, a (spin averaged) $T$-even contribution to the $\cos2\phi$
asymmetry
\bea
\nu_4=\frac{\frac{1}{Q^2}\sum_a e^2_a{\FF}
\left[w_4\,
f_1(x, k_\perp) \bar{f}_1(\bar{x}, p_\perp)
\right]}
{\sum_a e_a^2 {\FF}\left( f_1(x, k_\perp)
\bar{f}_1(\bar{x}, p_\perp)\right)},
%\nnn
\label{nu4}
\eea
where
%\bea
$w_4=2(\hat{\bm{h}}\cdot(\bm{k}_\perp -\bm{p}_\perp ))^2
-(\bm{k}_\perp-\bm{p}_\perp)^2$
%\nnn
%\eea
and, $\hat{\bm{h}}=\bm{q}_T/Q_T$. Indeed the earliest theoretical
explanation for azimuthal asymmetries was given by Collins and Soper's 
estimate of $\nu$ in Eq.~(\ref{nu4}), which is a kinematic non-leading
twist contribution.   However, all of the asymmetry functions, $\mu, \lambda$ and $\nu$,
were shown to  have parton model contributions.
Taking into account NLO~\cite{col79} and
NNLO~\cite{mirkes} the QCD improved parton model predicts
$1-\lambda-2\nu=0$, the so called  Lam-Tung relation~\cite{lam}.
However, experimental measurements of 
$\pi p \rightarrow \mu^+  \mu^- X$ discovered
unexpectedly large values of these asymmetries~\cite{e615,na10} compared to 
parton-model expectations resulting in a serious violation of this relation.
Attempts to account for the violation in terms of higher
twist effects~\cite{Berger-80,bran94} have been unsuccessful.  More recently,
 Boer~\cite{boerich} proposed that there is a dominant leading twist contribution to $\nu$ coming
from the $T$-odd transversity distributions $h_1^{\perp}(x,k_\perp)$ for
both hadrons which dominates in the kinematic range, $\bm{q_T}\ll Q$.
The $\cos 2\phi$ azimuthal asymmetry in  
unpolarized $p\, \bar{p}\rightarrow \mu^+\, \mu^-\, X$
would involve the convolution of the leading twist $T$-odd  function, $h_1^\perp$~\cite{boerich,bbh,ggmar,ma}
\bea
\nu_2 \hspace{-.25cm}&=&\hspace{-.25cm} \frac{\sum_a e^2_a {\FF}
\left[w_2\,
h_1^\perp(x, k_\perp)\bar{h}_1^\perp(\bar{x}, p_\perp)/
(M_1 M_2)\right]
}{\sum_a e_a^2 {\FF}
\left[f_1(x, k_\perp) \bar{f}_1(\bar{x}, p_\perp)\right]}
\nn
\label{nu2}
\eea
where $w_2=(2 \hat{\bm{h}}\cdot \bm{k}_{\perp }\cdot\hat{\bm{h}}\cdot
\bm{p}_{\perp }
- \bm{p}_\perp \cdot \bm{k}_\perp)$
is the weight in the convolution
integral, ${\FF}$~\cite{boerich}. A simple model for these distributions, inspired
by Collins' ansatz for the transversity fragmentation function led to
 $Q_T$  dependent $\nu$ which could
be fit to the low values of the $\pi p$ data.
Further work along those lines~\cite{bbh,ggmar,ma} incorporated a more
realistic model for the $T$-odd functions, as first developed
in SIDIS~\cite{bhs} for the functions $f_{1T}^{\perp}(x,k_\perp)$ which were
related to the $h_1^{\perp}(x,k_\perp)$ in ref.~\cite{gg}. 
The results were presented for $p\,  \bar{p}$
scattering.~\footnote{Very recently 
instanton induced effects have been investigated~\cite{boernach}.}  
This azimuthal asymmetry  is interesting in light of proposed experiments at Darmstadt  GSI~\cite{pax},
where an anti-proton beam is ideal for studying the
transversity property of quarks due to the dominance of {\em valence} quark
effects~\cite{anselpax}.  Herein we extend our calculations for $T$-odd contributions
to the unpolarized Drell-Yan Scattering first
reported in~\cite{ggmar}.  We perform a detailed analysis displaying 
$q_T$ and for the first time $x, x_F$, and  $q$ (or $m_{\mu\mu}$)   dependence of this
effect. In addition we compare  the double $T$-odd contribution to the conventional
subleading twist $T$-even contribution~\cite{cs}.    
\begin{figure}[htb]
\includegraphics[width=6.5cm]{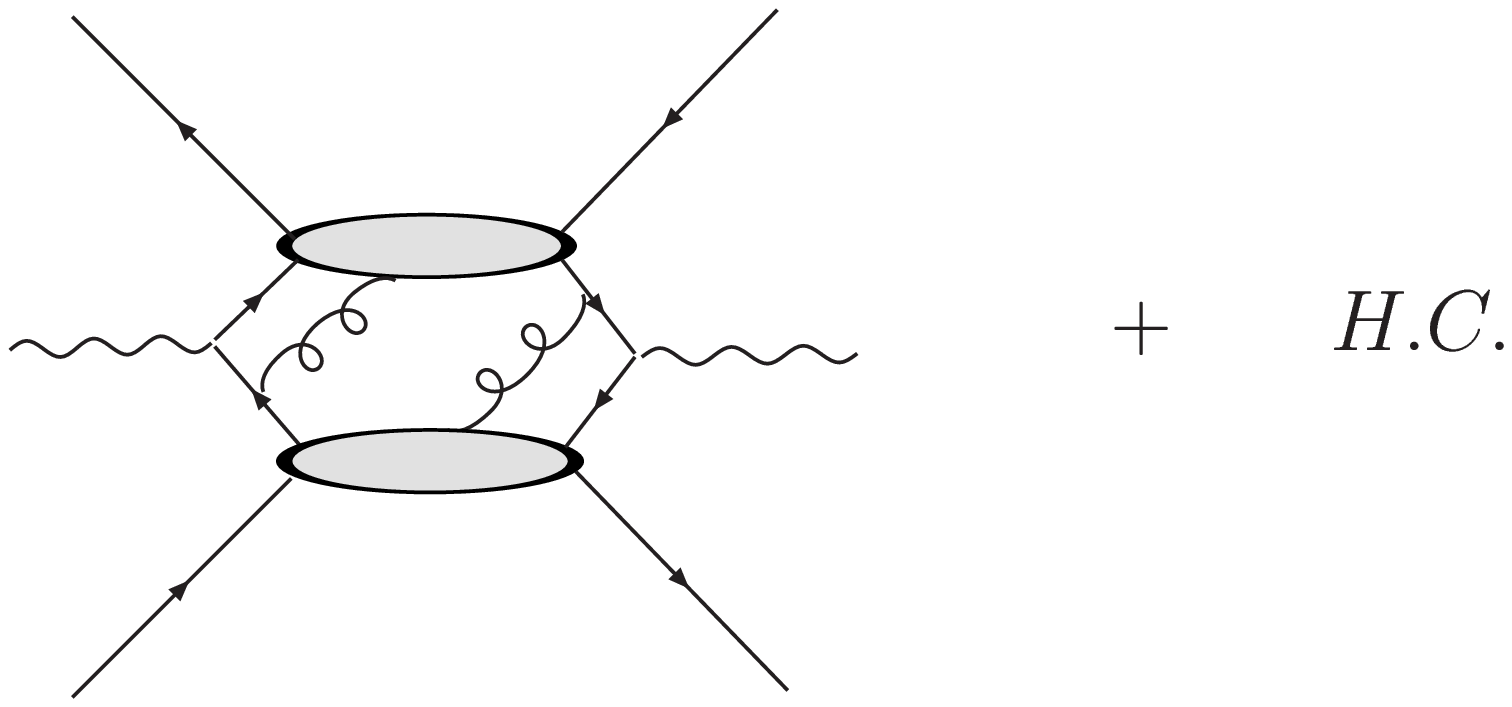}
\begin{center}
\includegraphics[width=5.0cm]{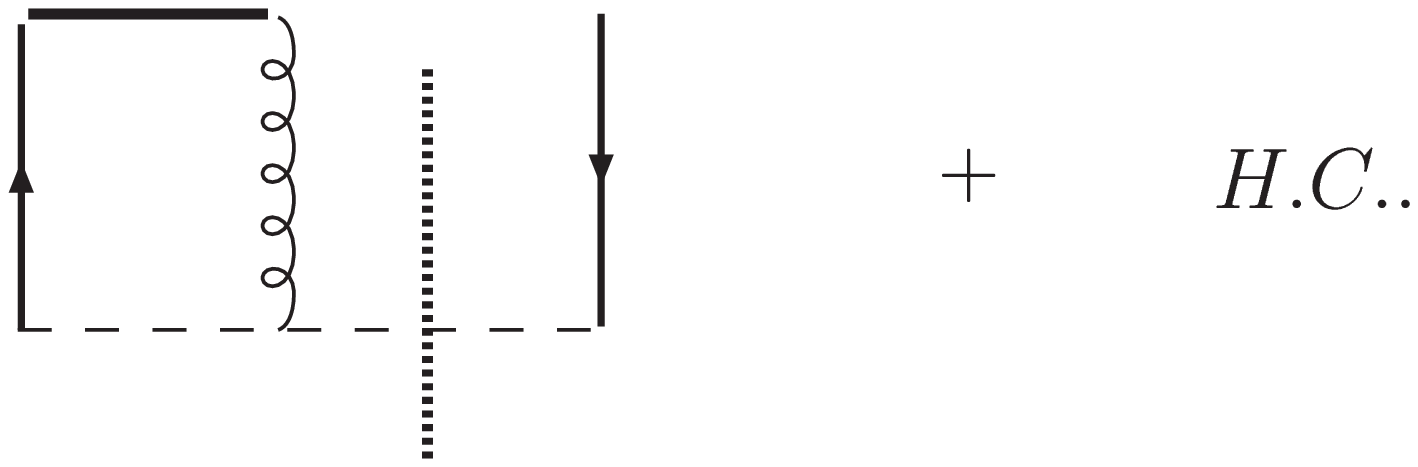}
\end{center}
\vspace*{-0.75cm}
\caption{\small {\em Above}: Feynman diagram representing
final state interactions giving rise
to $T$-odd contribution to
 Drell-Yan Scattering.
{\em Below}: Quark-target scattering amplitude
depicting  the $T$-odd contribution to the quark
distribution function in the eikonal approximation. }
\label{analyze}
\end{figure}

\section{$T$-odd transversity distribution}
For our purposes we consider $h_1^\perp$ projected from the correlation function for the
TMD  function $\Phi(k,P)$,
\bea
\Phi(x,\bm{k}_\perp)\hspace{-.35cm}&=&\hspace{-.35cm}
\frac{M}{2P^+}
\Big\{
f_1(x ,k_\perp) \frac{\slsh{P}}{M} 
\hspace{-.05cm}+\hspace{-.05cm} h_1^\perp (x,k_\perp)\frac{i\slsh{k}_\perp \slsh{P}}{M^2}
\nn && \hspace{3cm} + \dots \Big\}
\label{phi1}
\eea
that is
\bea
\int dk^-\Tr\left(\sigma^{\perp +}\gamma_5\Phi\right)\hspace{-.25cm}&=&\hspace{-.25cm}
%\nn && \hspace*{-2.5cm}\dots
\frac{2\varepsilon_{
\scriptscriptstyle
+-\perp j}k_{\scriptscriptstyle \perp j}}{M}h_1^\perp(x,k_\perp)
\dots\quad .
\nnn
\eea
In our work on SIDIS we used a parton model within the quark-diquark spectator
framework to model the quark-hadron interactions
that enter $\Phi(k,P)$~\cite{gg,ggoprh} and contribute to $T$-odd terms
in the projection. The basic diagram, indicating the one loop 
gluon exchange and the eikonalized
struck quark line arising from the gauge link is indicated in Figure~\ref{analyze}. 
Noting that parton intrinsic transverse 
momentum yielded a natural  regularization for the moments of these distributions,
we incorporated a Gaussian from factor  into our model~\cite{ggoprh,ggoprf}. The result was 
\bea
h_1^\perp(x,k_\perp)\hspace{-.25cm}&=&\hspace{-.25cm}
{\cal N}\alpha_s M
\frac{(1-x)(m+xM)}{k_\perp^2\Lambda(k^2_\perp)}{\cal R}_h(k_\perp^2;x)
\nn
\eea
where ${\cal R}$ is the regularization function
\bea
{\cal R}(k_\perp^2;x)\hspace{-.25cm}&=&\hspace{-.25cm}\exp^{-2b(k^2_\perp-
\Lambda(0))}
\nn  && \times
\left(\Gamma(0,2b \Lambda(0))-
\Gamma(0,2b \Lambda(k^2_\perp))\right).
\nnn
\eea
${\cal N}$ is a normalization factor determined 
with respect to  
the unpolarized $u$-quark distribution, obtained from the
zeroth moment of
\bea
f_1(x,k_\perp)\hspace{-.25cm}&=&\hspace{-.25cm}
\frac{{\cal N}(1-x)\left(\left(m\hspace{-0.05cm}+\hspace{-0.05cm}xM\right)^2
\hspace{-0.05cm}+\hspace{-0.05cm}k_\perp^2\right)}
{\Lambda^2(k_\perp^2)}{\cal R}_f(k_\perp^2),
\nnn
\eea
normalized with respect to valence distributions in~\cite{GRV}.  
${\cal R}_f(k_\perp^2;x)=\exp^{-2bk_\perp^2}$.

The asymmetry in the $p\bar{p}$ Drell-Yan process involves the
convolution of the product of two $T$-odd distributions.
The  contribution to the double $T$-odd azimuthal
$\cos 2\phi$ asymmetry, $\nu$ in Eq.~(\ref{nu2}),
in terms of initial and final (ISI/FSI)
state interactions of active or ``struck'', and  fragmenting
quark~\cite{ggoprh,ggoprf}
is depicted in Figure~\ref{analyze}.
\begin{figure}[htb]
%\begin{center}
\centerline{\epsfxsize=7.5cm \epsfbox{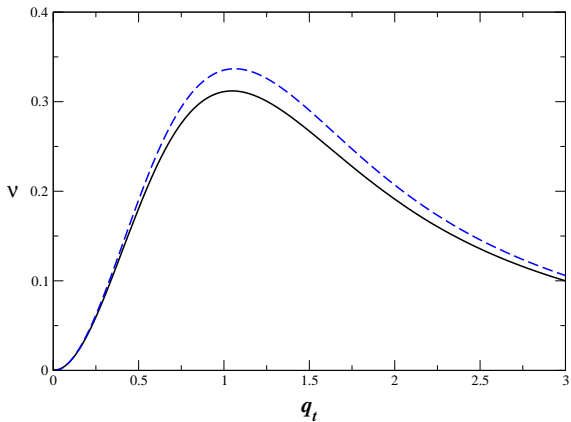} }
\vspace*{-1cm}
\caption{\small $\nu$ plotted as a function of $q_T$ for
$s=50\ {\rm GeV}^2$, $x$ in the range $0.2-1.0$, and  $q$ ranging from
$3-6 \ {\rm GeV}/c$.}
\label{nu_qt}
%\end{center}
\end{figure}
%\vskip 1cm
\begin{figure}[htb]
\vspace*{0.25cm}
%\begin{center}
\centerline{\epsfxsize=7.5cm \epsfbox{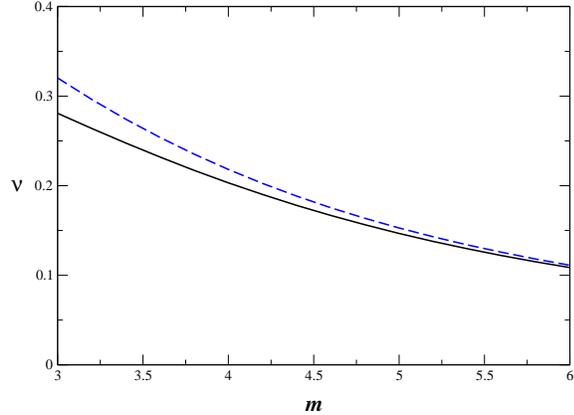} }
\vspace*{-1cm}
\caption{\small $\nu$ plotted as a function of $q=m_{\mu\mu}$ for
$s=50\ {\rm GeV}^2$, $x$ in the range $0.2-1.0$, and  $q_T$ ranging from
$0-3 \ {\rm GeV}/c$.}
\label{nu_q}
%\end{center}
\end{figure}
\begin{figure}[htb]
\vspace*{0.4cm}
%\begin{center}
\centerline{\epsfxsize=7.5cm \epsfbox{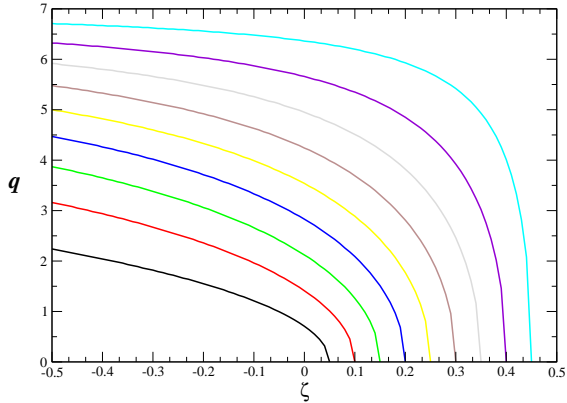} }
\vspace*{-1cm}
\caption{\small Contours of constant 
$x$ as a function of $\zeta$ and $q$  which  ranges from 0 to 3 GeV/c.}
\label{zeta}
%\end{center}
\end{figure}
\begin{figure}[htb]
\vspace*{0.5cm}
%\begin{center}
\centerline{\epsfxsize=7.0cm \epsfbox{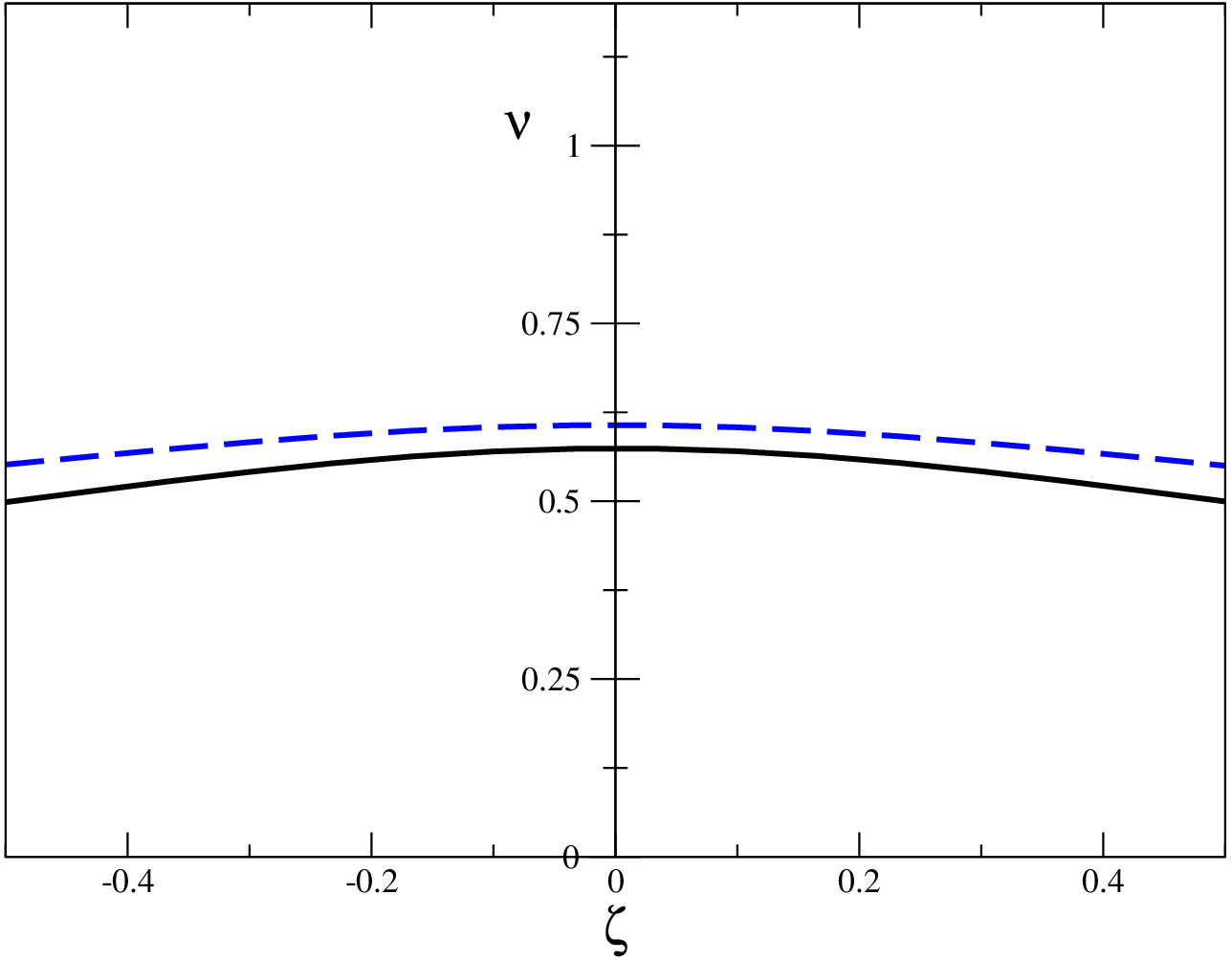} }
\vspace*{-1cm}
\caption{\small $\nu$ plotted as a function of $\zeta$ for
$s=50\ {\rm GeV}^2$, $q_T$ ranging from $3$ to $6$ GeV/c and $q$ from
0 to 3 GeV/c.}
\label{nuzeta}
%\end{center}
\end{figure}
\begin{figure}[htb]
\vspace*{0.25cm}
%\begin{center}
\centerline{\epsfxsize=7.5cm \epsfbox{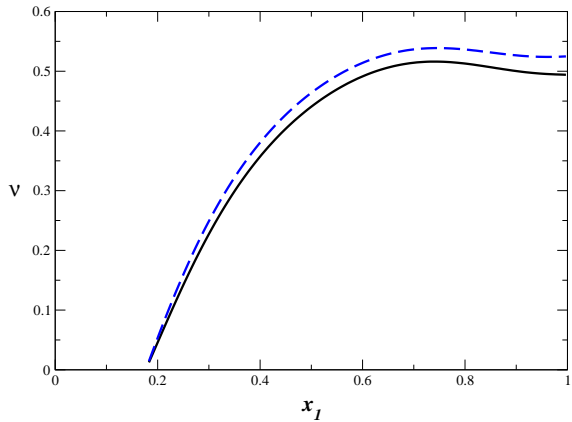} }
\vspace*{-1cm}
\caption{\small $\nu$ plotted as a function of $x$ for
$s=50\ {\rm GeV}^2$ $q_T$ ranging from 3 to 6 GeV/c and $q$ from
0 to 3 GeV/c.}
\label{nux}
%\end{center}
\end{figure}
\begin{figure}[htb]
\vspace*{0.30cm}
%\begin{center}
\centerline{\epsfxsize=7.0cm \epsfbox{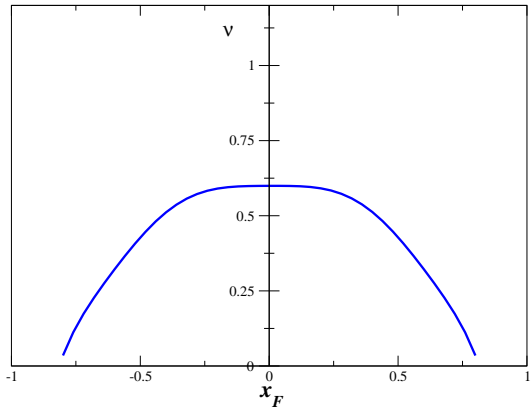} }
\vspace*{-1cm}
\caption{\small The leading twist contribution to $\nu$ plotted as a function of $x_F$ for
$s=50\ {\rm GeV}^2$ $q_T$ ranging from 3 to 6 GeV/c and $q$ from
0 to 3 GeV/c.}
\label{nuxf}
%\end{center}
\end{figure}
We have numerically performed the convolution integrals and 
obtained values of the asymmetry $\nu$ as a function of
the variables, $x$, $x_F$, $q_T$, and $q$ (or $m_{\mu\mu}$).

Before evaluating the convolution, the Drell-Yan kinematics demand some
special attention. With $x$ and $\bar x$  being the
fractional longitudinal momenta of the quark and antiquark, there are some
constraints:
\bea
x\bar{x}\hspace{-.25cm}&=&\hspace{-.25cm}
\tau=Q^2/s\quad , \quad \frac{x-\bar{x}}{2}\equiv\eta=x_F/2\quad {\rm and}
\nn &&\hspace{-1.1cm}  
x=\eta+\sqrt{\eta^2+\tau^2}\, ,\quad
\bar{x}=-\eta+\sqrt{\eta^2+\tau^2}.
\eea
Due to the constraint
$x\bar{x} = q^2/s$ the allowed range of $x$ is restricted for each
$q$ value, from $x_{min}=q^2/s$ to 1. Furthermore, evaluating the convolutions of
$h_1^\perp \bar{h}_1^\perp$ and $f_1 \bar{f}_1$ for a sampling of $x$
will not treat the $\bar{x}$ and the corresponding antiparticle structure
functions symmetrically. So it is more appropriate to use the symmetrical
variable, Feynman-$x$, $x_F = x - \bar{x}$ or $x =\frac{1}{2}(x_F+\sqrt{x_F^2+4q^2/s})$
and $\bar{x} =\frac{1}{2}(-x_F+\sqrt{x_F^2+4q^2/s})$. However, the allowed range
of $x_F$ depends on $q$ (from $-1/2 (1-q^2/s)$ to $+1/2 (1-q^2/s)$)--the
variables $x_F$ and $q$ are not orthogonal. Since we aim to present
partially integrated values of $\nu$, approximating experimentalists'
measurements, it is advantageous to work with orthogonal
variables. We choose the variable
\bea
\zeta = \frac{1}{2}\frac{x_F}{(1-q^2/s)},
\eea
with range from $-\frac{1}{2}$ to $+\frac{1}{2}$, independent of $q$.
Values of the asymmetry fill the
rectangular space of variables, $\zeta, q, q_T$. We have been careful with
this choice because our model predictions have considerable structure in all
3 variables. Hence the meaning of a graph of $\nu(q)$ or $\nu(x)$
has particular significance when comparing to experimental data.

A crucial point in selecting these variables involves how experimenters
determine various asymmetries and angular dependences, in order to maximize
statistics when extracting possibly small effects like $\nu$. Events appear
distributed over allowed regions (modified by experimental acceptances) of
all three variables along with the $\mu$ pair angular variables, of course.
To obtain the dependence on one variable, large bins are defined and event
numbers averaged over those bins. How are those results to be compared with
theoretical predictions~\cite{radici}\ ?  The two experiments for which data have been
published have different ranges of variables~\cite{e615,na10}. The binning
procedures are not easily compared. To be most general and adaptable for
future experimental comparisons we have determined the value of $\nu$ as a
function of $\zeta, q, q_T$. We  then integrate over pairs of those variables
for particular ranges of the variables. At $s =50\ {\rm GeV}^2$ we
take $q_T\le 3\ {\rm GeV}/c$ and $3\ {\rm GeV}/c \le q \le 6 {\rm GeV}/c$, while $\zeta$
always varies from -1/2 to +1/2.

For $\nu(q_T)$ and $\nu(q)$ the resulting values are shown in
Figures~\ref{nu_qt} and \ref{nu_q}.
There is a corresponding $\nu(\zeta)$ shown in Figure~\ref{nuzeta}.
To connect with the $x$ or $x_F$ dependence we have to take the $q$ dependence
into account. For any $q_T$ the fixed $x_F$ values form contours in
the $\zeta, q$ plane (see Figure~\ref{zeta}). So an integral of $\nu$ over
$q$ for a fixed $x_F$ follows the relevant contour in $\zeta, q$ and has a
limit on $\tau=q^2/s$ of $(1-\frac{x_F}{2\zeta})$. This limits the range
of the $q$ integral until the limiting value of the range at $q = 6\ {\rm GeV}/c$
for $s = 50\  {\rm GeV}^2$. Similarly, for any  $q_T$ the fixed $x$ values form asymmetrical contours
in the $\zeta, q$ plane as shown in Figure~\ref{zeta}. 
The limit on $\tau$ for a
fixed $x$ will be $x(\frac{2\zeta -x}{2\zeta x-1})$. 
The resulting values
of $\nu(x)$ are shown in Figure~\ref{nux}. Finally, in Figure~\ref{nuxf} we plot the leading 
twist contribution to $\nu$ as a function
of $x_F$.

\section{$T$-Even Contribution}

Long before the realization that there is a leading twist $2$ contribution to the Drell-Yan azimuthal asymmetry, 
it was proposed by Collins and Soper~\cite{cs} that the spin independent, transverse momentum dependent 
distributions $f_1$ and $\bar{f}_1$ could contribute via Eq.~(\ref{nu4}). It is important to compare this 
{\em kinematic} twist $4$ contribution to the leading twist contribution (Eq.~(\ref{nu2}) shown above. We combined both 
convolutions to determine the magnitude of the shift. The additional contribution for $s=50~{\rm GeV}^2$ to each 
of the partially integrated functions $\nu$ is shown in Figures~\ref{nu_qt},\ref{nu_q},\ref{nuzeta},\ref{nux}
as slightly higher curves. At most the additional contribution is around 3--4\%. For higher $s$ values the effect 
is even smaller, as expected~\cite{ggmar}.

\section{Conclusion}

A perusal of the figures shows that the $\cos 2\phi$ azimuthal asymmetry $\nu$ 
is not small at center of mass energies of $50\ {\rm GeV}^2$.
We estimated the leading twist $2$ and twist $4$ contributions~\cite{ggmar}.
In Figure~\ref{nu_qt}, the $T$-odd portion contributes about $30\% $ with an
additional $3\%$ from the
sub-leading  $T$-even piece.
The distinction between
the leading order $T$-odd and  sub-leading order
$T$-even contributions diminish
at center of mass energy of $s=500\ {\rm GeV}^2$~\cite{ggmar}.
In  Figure~\ref{nux}, $\nu$ is plotted versus $x$ at $s=50\ {\rm GeV}^2$,
where $q_T$ ranges from 2 to 4 GeV.  Again the higher twist contribution is small.

Thus, aside from the
competing $T$-even effect,  the experimental
observation of a strong $x$-dependence
would indicate the presence of $T$-odd structures in {\it unpolarized} Drell-Yan scattering, 
implying that novel transversity properties of the  nucleon can be accessed  {\em without invoking beam or 
target polarization}.

It should be noted that at order $\alpha_s$ a complete analysis for the full range 
of $q_T$ would entail including gluon bremsstrahlung
contributions~\cite{col79}. Furthermore, collinear Sudakhov corrections have not been accounted for 
here~\cite{boersud}. 
A thorough explication of Drell-Yan dynamics would require more care with regions in 
which divergent contributions become important to address. For this study, however, we have considered the 
implications of our model, unencumbered by subtleties at the edges of the phase space on which we concentrate.

We conclude that $T$-odd  correlations of intrinsic transverse
quark momentum and transverse spin of quarks  are intimately
connected with studies of the $\cos 2\phi$
azimuthal asymmetries in $p\bar{p}$--Drell-Yan scattering.
Due to the dominance of valence quark effects 
% in $p\bar{p}$--Drell-Yan scattering
we estimate that the proposed proton anti-proton
experiments at GSI~\cite{pax} provide an excellent opportunity 
to study the role that  $T$-odd  correlations
play in characterizing intrinsic transverse spin effects
within the proton.

\begin{flushleft}
{\em Acknowledgement}
\end{flushleft}
\vspace*{-0.2cm}
We thank Daniel Boer and Karo Oganessyan  for useful discussions.   
GG acknowledges partial support from U.S. Department of Energy grant
DE-FG02-92ER40702.

\end{document}